\documentclass[prd, amsmath,superscriptaddress,nofootinbib]{revtex4-2}
\usepackage{graphicx}
\usepackage{bm}
\usepackage{color}
\usepackage{amsmath,amssymb}	
\usepackage{setspace}
\usepackage{hyperref}
\usepackage{longtable}
\usepackage{booktabs}
\usepackage{multirow}

\usepackage{slashed}
\usepackage{comment}
\usepackage{ulem}
\usepackage{amsmath,amsfonts,amssymb}
\usepackage{ifthen}

\usepackage{physics}
\begin{document}

\def\uphi{\chi}
\def\suphi{\textcolor{red}{\phi}}
\newcommand{\monohiggs}{\phi}
\newcommand{\strhiggs}{\eta}
\def\mD{\mathrm{D}}
\def\tM{\theta_{\mathrm{mix}}}
\def\a{\alpha}
\def\b{\beta}
\def\c{\varepsilon}
\def\d{\delta}
\def\e{\epsilonsilon}
\def\f{\phi}
\def\g{\gamma}
\def\h{\theta}
\def\k{\kappa}
\def\l{\lambda}
\def\m{\mu}
\def\n{\nu}
\def\p{\psi}
\def\q{\partial}
\def\r{\rho}
\def\s{\sigma}
\def\t{\tau}
\def\u{\upsilon}
\def\v{\varphi}
\def\w{\omega}
\def\x{\xi}
\def\y{\eta}
\def\z{\zeta}
\def\D{\Delta}
\def\G{\Gamma}
\def\H{\Theta}
\def\L{\Lambdabda}
\def\F{\Phi}
\def\P{\Psi}
\def\S{\Sigma}

\def\o{\over}
\def\beq{\begin{eqnarray}}
\def\eeq{\end{eqnarray}}
\newcommand{\gsim}{ \mathop{}_{\textstyle \sim}^{\textstyle >} }
\newcommand{\lsim}{ \mathop{}_{\textstyle \sim}^{\textstyle <} }
\newcommand{\EV}{ {\rm eV} }
\newcommand{\KEV}{ {\rm keV} }
\newcommand{\MEV}{ {\rm MeV} }
\newcommand{\GEV}{ {\rm GeV} }
\newcommand{\TEV}{ {\rm TeV} }
\newcommand{\1}{\mbox{1}\hspace{-0.25em}\mbox{l}}
\newcommand{\headline}[1]{\noindent{\bf #1}}
\def\diag{\mathop{\rm diag}\nolimits}
\def\Spin{\mathop{\rm Spin}}
\def\SO{\mathop{\rm SO}}
\def\O{\mathop{\rm O}}
\def\SU{\mathop{\rm SU}}
\def\U{\mathop{\rm U}}
\def\Sp{\mathop{\rm Sp}}
\def\SL{\mathop{\rm SL}}
\def\tr{\mathop{\rm tr}}
\def\mpl{M_{PL}}

\def\IJMP{Int.~J.~Mod.~Phys. }
\def\MPL{Mod.~Phys.~Lett. }
\def\NP{Nucl.~Phys. }
\def\PL{Phys.~Lett. }
\def\PR{Phys.~Rev. }
\def\PRL{Phys.~Rev.~Lett. }
\def\PTP{Prog.~Theor.~Phys. }
\def\ZP{Z.~Phys. }

\def\ff{\mathrm{f}}
\def\BH{{\rm BH}}
\def\inf{{\rm inf}}
\def\ev{{\rm evap}}
\def\eq{{\rm eq}}
\def\SM{{\rm sm}}
\def\Mpl{M_{\rm PL}}
\def\GeV{{\rm GeV}}
\newcommand{\version}{arXivv2}
\newcommand{\modifiedat}[2]{%
\ifthenelse{\equal{\version}{#1}}{\textcolor{red}{#2}}{#2}%
}
\newcommand{\Red}[1]{\textcolor{red}{#1}}

\newenvironment{draftpart}{\vspace{1cm}\hrule\relax\textbf{DRAFT FROM HERE}\par}{\textbf{DRAFT UNTIL HERE}\hrule\relax\vspace{1cm}}
\newcommand{\rqed}{\mathrm{QED}}
\newcommand{\uone}[1][\rqed]{\mathrm{U(1)}_{#1}}
\newcommand{\eqrefer}[1]{Eq.\,\eqref{#1}}
\newcommand{\matform}[1]{\mathcal{#1}}
\newcommand{\ncharge}[1][\rqed]{{n^{e}_{#1}}}
\newcommand{\nchargep}[1][\rqed]{{n^{e\,\prime}_{#1}}}
\newcommand{\qcharge}[1][\rqed]{{Q^{e}_{#1}}}
\newcommand{\suD}{\mathrm{SU}(2)_\mD}
\newcommand{\lagrangian}{\mathcal{L}}
\newcommand{\azimuth}{\varphi_A}
\newcommand{\zenith}{\theta_Z}

\title{Interactions of electrical and magnetic charges and dark topological defects}

\author{Akifumi Chitose}
\email[e-mail: ]{achitose@icrr.u-tokyo.ac.jp}
\affiliation{Institute for Cosmic Ray Research (ICRR), The University of Tokyo, Kashiwa, Chiba 277-8582, Japan}
\author{Masahiro Ibe}
\email[e-mail: ]{ibe@icrr.u-tokyo.ac.jp}
\affiliation{Institute for Cosmic Ray Research (ICRR), The University of Tokyo, Kashiwa, Chiba 277-8582, Japan}

\date{\today}

\begin{abstract}
We consider a model of dark photon which appears as a result of the successive symmetry breaking SU(2)$\,\to\,$U(1)$\,\to \mathbb{Z}_2$, where various types of topological defects appear in the dark sector.
In this paper, we study the interactions between QED charges and the dark topological defects through mixing between QED photon and dark photon. 
In particular, we extend our previous analysis by incorporating the magnetic mixing and $\theta$-terms. 
We also consider the
dyons and dyonic beads in the dark sector.
Notably, dark magnetic/dyonic beads are found to induce a QED Coulomb potential through the magnetic mixing
despite finite mass of the dark photon.
\end{abstract}

\maketitle

\section{Introduction}
The dark photon~\cite{Holdom:1985ag}, a massive vector boson
which slightly mixes with the QED photon, 
appears in various extensions of the Standard Model (SM).
Recently, applications of dark photons to cosmology have been actively discussed. For instance, sub-GeV dark photons can mediate dark matter self-interactions, possibly providing 
a better fit to the small scale structure of the Universe~\cite{Spergel:1999mh,Kaplinghat:2015aga,Kamada:2016euw,Tulin:2017ara,Chu:2018fzy,Chu:2019awd}.
The dark photon may also play an essential role in sub-GeV dark matter models, as it can transfer excess entropy in the dark sector to the SM sector before the neutrino decoupling
(see e.g. Ref.~\cite{Blennow:2012de,Ibe:2019gpv}).
Following the attention, sub-GeV dark photons have been an important search target for various experiments (see e.g., Refs.~\cite{Raggi:2015yfk,Bauer:2018onh} for the current experimental status).

More plausible dark photon scenarios require more serious discussions of the origin of the dark photon mass.
One possibility is to identify the dark photon model with the St\"u\modifiedat{PRDv2}{c}kelberg model (see Ref.~\cite{Ruegg:2003ps} for a review).
As the model requires no new particles other than the massive vector boson, it provides the simplest model of the dark photon.
However, such a model is shown to violate unitarity~\cite{Kribs:2022gri}.%
\footnote{\modifiedat{PRDv3}{Although the interaction of St\"uckelberg
vector boson and a conserved current does not violate unitarity, other interactions such as self-couplings do.}}
Thus, it seems more compelling to assume that the dark photon mass originates from spontaneous U(1) symmetry breaking.%
\footnote{In addition to the conventional Higgs mechanism, it is also possible to break the U(1) gauge symmetry dynamically~\cite{Co:2016akw,Ibe:2021gil}.}

Once we assume spontaneous U(1) breaking
in the dark sector, 
its extension to non-Abelian gauge theory would be of interest.
Aside from purely theoretical interest, potential high energy asymptotic freedom motivates such extensions as a UV completion of the U(1) model.
It is also attractive as it can naturally explain tiny mixing parameters (see e.g., Refs.~\cite{Ibe:2018tex,Ibe:2019ena}).
The smallness of the mixing parameters is important to evade all the astrophysical, cosmological, and experimental constraints.

In Ref.~\cite{Hiramatsu:2021kvu}, it has been discussed how topological defects
in the dark sector affect the SM sector
through the kinetic mixing 
when the the dark photon 
originates from an SU(2) gauge symmetry.
In this setup, various topological defects appear, including magnetic monopoles, strings, and magnetic beads.
In particular, Ref.~\cite{Hiramatsu:2021kvu} showed that dark magnetic beads induce a configuration that looks like a QED magnetic monopole from a distance through kinetic mixing, while retaining the QED Bianchi identity.

In this paper, we extend the analysis of Ref.~\cite{Hiramatsu:2021kvu}
by adding the magnetic mixing term~\cite{Brummer:2009oul}
between the dark photon and the QED photon.
We also discuss how dyons (and the dyonic beads) in the dark sector 
affect QED configurations.
Charge quantization in the presence of the mixing terms and the $\theta$-term is also considered.

In our analysis, (and the analysis in Ref.~\cite{Hiramatsu:2021kvu}),
we explicitly discuss SU(2) gauge theory 
behind the topological defects such as monopoles and dyons,
which clarifies how and when the $\theta$-terms as well as the magnetic mixing become effective.
This approach provides a complementary 
understanding to the previous studies 
in Refs.~\cite{Brummer:2009cs,Brummer:2009oul,Long:2014mxa,Terning:2018lsv,Terning:2019bhg,Hook:2017vyc}
on how the dark monopoles/strings affect the QED sector 
through the mixing within the effective U(1) theory.

The organization of the paper is as follows.
In Sec.~\ref{sec:Setup}, we summarize our setup where the dark photon appears
from a successive symmetry breaking SU(2)$\,\to\,$U(1)$\,\to\mathbb{Z}_2$.
In Sec.~\ref{sec:U1symmetric} and Sec.~\ref{sec:U1broken},
we discuss the QED interactions of dark charged objects through 
the kinetic and magnetic mixing in the $\uone[]$ symmetric and broken phases, respectively.
The final section is devoted to our conclusions.

\section{Dark Photon from Non-Abelian Gauge Theory}
\label{sec:Setup}
In this paper, we discuss 
the effects of charged
objects in the dark sector
including topological defects such as monopoles/dyons/strings/beads
which are expected to appear 
in the successive symmetry breaking,
SU(2)$\,\to\,$U(1)$\,\to \mathbb{Z}_2$.
Hereafter, we call these gauge groups
$\mathrm{SU}(2)_\mathrm{D}$ and  $\mathrm{U}(1)_\mathrm{D}$, respectively.

\subsection{
U(1)\texorpdfstring{${}_\mathrm{QED}\times$}{QED x}
SU(2)\texorpdfstring{${}_\mD$}{D} Model}
\label{sec:U1SU2}
We consider a $\mathrm{U}(1)_\mathrm{QED}\times \mathrm{SU}(2)_\mD$ gauge theory where the 
two sectors are coupled through higher dimensional operators:%
\footnote{We take the spacetime metric as $(g_{\mu\nu})=(-1,1,1,1)$.}
\begin{align}
\label{eq:SU2model}
\mathcal{L}&=-\frac{1}{4}F_{\mu\nu}F^{\mu\nu}-\frac{1}{4}{F^a_\mD}_{\mu\nu}F_\mD^{a\mu\nu}
-\frac{1}{2}D_{\mu}\monohiggs^a D^{\mu}\monohiggs^a
-\frac{1}{2}D_{\mu}\strhiggs^a D^{\mu}\strhiggs^a
-V(\monohiggs,\strhiggs) + \mathcal{L}_\theta + \mathcal{L}_{\mathrm{mix}}, \\
\label{eq:Ltheta}
\mathcal{L}_\theta &= -\frac{e^2\theta}{32\pi^2}F_{\mu\nu}\tilde{F}^{\mu\nu} - \frac{e_\mD^2\theta_\mD}{32\pi^2}F_{\mD\mu\nu}^a\tilde{F}_\mD^{a\mu\nu} \\
\label{eq:Lmix}
\mathcal{L}_{\mathrm{mix}} &= -\frac{c_1\monohiggs^a}{2\Lambda} {F_\mD}^{a}_{\mu\nu}F^{\mu\nu}-
  \frac{c_2\monohiggs^a}{16\pi^2\Lambda} {F_\mD}^{a}_{\mu\nu}\tilde{F}^{\mu\nu}\ .
\end{align}
Here, $F_{\mu\nu}$ and $F^a_{\mD\mu\nu}$
($a = 1,2,3$)
are the field strengths of the $\uone$ and SU$(2)_\mD$ gauge fields, $A_{\mu}$
and $A^a_{\mD\mu}$, respectively.
Their hodge duals are given by $\tilde{F}_{(\mD)\mu\nu} = \epsilon_{\mu\nu\rho\sigma}F_{(\mD)}^{\rho\sigma}/2$.\footnote{We adopt the convention $\epsilon_{0123} = 1$.
The three dimensional anti-symmetric tensor is $\epsilon_{ijk}=\epsilon^{ijk}=\epsilon_{0ijk}=\epsilon_0{}^{ijk}$.
We also define electromagnetic fields as
$E^i = F^{0i}=F_{i0}$ and 
$B^i = \epsilon^{ijk}F^{jk}/2$.
}
We introduced two SU$(2)_\mD$ adjoint scalar fields $\monohiggs^{a}$ and $\strhiggs^a$ ($a = 1,2,3$).
We call the $\uone$ and $\suD$ gauge coupling constants $e$ and $e_\mD$.
The covariant derivatives of $\monohiggs$ and $\strhiggs$ are given by,
\begin{align}
    D_\mu \monohiggs^a &= \partial_\mu \monohiggs^a + e_\mD \epsilon^{abc}A_{\mathrm{D}\mu}^b\monohiggs^c \ , \\
    D_\mu \strhiggs^a &= \partial_\mu \strhiggs^a + e_\mD \epsilon^{abc}A_{\mathrm{D}\mu}^b\strhiggs^c\ .
\end{align} 
The higher dimensional operators 
with coefficients $c_{1,2}$ suppressed by the UV cutoff $\Lambda$ result in 
effective mixing parameters between QED photons and dark photons~\cite{Brummer:2009oul}.
We take $\Lambda \gg v_1$, so that the effective mixing parameters are small.
Throughout this paper, we assume that no SU(2)$_\mD$ charged fields have U(1)$_\mathrm{QED}$ charge,
although our discussion can be generalized.

The scalar potential of $\monohiggs^a$ and $\strhiggs^a$ is assumed to be
\begin{equation}
\label{eq:potential}
	V(\monohiggs,\strhiggs) = \frac{\lambda_1}{4}
	(\monohiggs\cdot\monohiggs-v_1^2)^2 + \frac{\lambda_2}{4}(\strhiggs\cdot\strhiggs-v_2^2)^2 + \frac{\kappa}{2}(\monohiggs\cdot \strhiggs)^2 \ ,
\end{equation}
where $\monohiggs\cdot\monohiggs= \monohiggs^a\monohiggs^a$ etc.
For simplicity, we omit terms such as $(\monohiggs\cdot\monohiggs)(\strhiggs\cdot\strhiggs)$.
The dimensionless coupling constants $\lambda_1, \lambda_2$ and $\kappa$ are taken to be positive.
We also take the 
mass scales to be hierarchical, i.e., $v_1 \gg v_2$.
At the vacuum, $\monohiggs^a$ takes the trivial configuration, 
\modifiedat{PRDv2}{i.e. the vacuum expectation value (VEV),}
\begin{equation}
\label{eq:VEV1}
   \langle{\monohiggs^a}\rangle = v_1 \delta^{a3}\ ,
\end{equation}
with which $\mathrm{SU}(2)_\mD$ is broken down to $\mathrm{U}(1)_\mD$.
The remaining $\mathrm{U}(1)_\mD$ symmetry corresponds to the SO(2) symmetry around the $a=3$ axis of $\mathrm{SO}(3)\simeq\mathrm{SU}(2)_\mD$ vectors $\monohiggs^a$ and $\strhiggs^a$.

Below the $\suD$ breaking scale,
a U(1)$_\mD$ charged field $\uphi$ can be formed out of $\strhiggs^a$ as
\begin{equation}\label{eq:u1phi}
    \uphi = \frac{1}{\sqrt{2}}(\strhiggs^1-i\strhiggs^2)\ .
\end{equation}
For $\kappa>0$, 
the last term of the potential \eqref{eq:potential} 
lifts the $a=3$ component of $\strhiggs$ and the VEV of $\strhiggs^a$ is required to be orthogonal to 
$\langle\monohiggs^a\rangle$. 
As a result, $\langle\strhiggs^a\rangle$ takes a value in the $(\strhiggs^1,\strhiggs^2)$ plane, i.e.,
\begin{equation}\label{eq:VEV2}
    \langle\strhiggs^a\rangle = v_2 \delta^{a1}\ , 
\end{equation}
or 
\begin{equation}
    \label{eq:U1VEV}
    \langle\uphi\rangle = \frac{1}{\sqrt{2}} v_2\
     ,
\end{equation}
which breaks $\mathrm{U}(1)_\mD$ spontaneously.
In this way, successive symmetry breaking
$\mathrm{SU}(2)_\mD\to \mathrm{U}(1)_\mD \to \mathbb{Z}_2$
is achieved.
The $\mathbb{Z}_2$ 
symmetry is the center of $\mathrm{SU}(2)_\mD$.

\subsection{Effective 
U(1)\texorpdfstring{${}_\mathrm{QED}\times$}{QED x}
U(1)\texorpdfstring{${}_\mD$}{D} Theory}
\label{ssec:U1model}
For later use, 
we describe the effective 
U(1)\texorpdfstring{${}_\mathrm{QED}\times$}{}
U(1)\texorpdfstring{${}_\mD$}{} theory
for $\expval{\monohiggs} \ne 0$.
The effective Lagrangian is given by
\begin{align}
\label{eq:U1model}
\mathcal{L}=&-\frac{1}{4}F_{\mu\nu}F^{\mu\nu}-\frac{1}{4}{F_\mD}_{\mu\nu}F_\mD^{\mu\nu}+\frac{\epsilon}{2}F_{\mu\nu}F_\mD^{\mu\nu}
-\frac{\tM}{16\pi^2}F_{\mu\nu}\tilde{F}_{\mD}{}^{\mu\nu}
- \frac{e^2\theta}{32\pi^2}F_{\mu\nu}
\tilde{F}^{\mu\nu} - \frac{e_\mD^2\theta_\mD}{32\pi^2}F_{\mD\mu\nu}
\tilde{F}_\mD^{\mu\nu}
\cr
&+e A_\mu J_\mathrm{QED}^\mu 
+ e_{\mD} A_{\mD\mu} J_\mD^\mu
-D_{\mu}\uphi D^{\mu}\uphi^*-V(\uphi) \ , 
\end{align}
where $F_{\mD\mu\nu} = \monohiggs^a F^a_{\mD\mu\nu} / v_1$ represents the $\uone[\mD]$ gauge field strength and $A_{\mD\mu}$ the corresponding gauge field.
We call the gauge field $A_{\mD\mu}$ as the dark photon.
Note that in the presence of monopoles/dyons, the effective theory
is well-defined only far enough from them (so that $\abs{\monohiggs} = v_1$)
and $A_{\mD\mu}$ can be defined only locally.
We also explicitly displayed the currents $J_{\mathrm{QED}}^\mu$ and $J_\mD^\mu$ coupled to the gauge fields, which were omitted in Eq.\,\eqref{eq:SU2model}.

We refer to the interactions with the couplings $\epsilon$ and $\tM$ as the kinetic and magnetic mixing.
They arise from the higher dimensional operators \eqref{eq:Lmix},
where 
the couplings are related to the underlying model parameters by
\begin{equation}
    \epsilon = \frac{c_1v_1}{\Lambda}\ , \quad 
    \tM = \frac{c_2v_1}{\Lambda}\ .
\end{equation}
As we assume $\Lambda \gg v_1$, these parameters are tiny.\footnote{
The parameter $\tM$
is related to $\theta_{12}$ in Ref.~\cite{Brummer:2009oul} via $\theta_{12}=ee_\mD\tM$.
}

In the effective U$(1)_\mD$ theory,
only $\uphi$
is relevant as the other components become heavy 
for $\kappa>0$.
The covariant derivative of $\uphi$ is given by
\begin{equation}
D_{\mu}\uphi=(\partial_{\mu}-ie_\mD A_{\mD\mu})\uphi\ . 
\end{equation}
The scalar potential $V(\uphi)$ is obtained by substituting Eqs.\,\eqref{eq:VEV1}, \eqref{eq:u1phi}, and \eqref{eq:VEV2} into \eqrefer{eq:potential}:
\begin{equation}
V(\uphi)=\frac{\lambda}{4}(|\uphi^2|-v^2)^2\ ,
\end{equation}
where $\lambda = \lambda_2/2$ and $v=v_2/\sqrt{2}$.
At the vacuum, 
$\uphi$ obtains a VEV $\langle \uphi \rangle = v$, which spontaneously breaks the $\mathrm{U}(1)_\mD$ symmetry, as in the previous subsection.

\section{\texorpdfstring{$\mathrm{U(1)}_\mD$}{U(1)D} Symmetric Phase}
\label{sec:U1symmetric}
In this section, we discuss 
the effects of electrically and magnetically charged
objects in the dark sector
in the U(1)$_\mD$ symmetric phase
by ignoring $\strhiggs$.

\subsection{Dark Elementary Charged Particles}

Let us consider the effective U$(1)_\mathrm{QED}\times$U$(1)_\mD$ theory \eqref{eq:U1model} assuming the 
trivial vacuum \eqref{eq:VEV1} with charged particles in $J_{\mathrm{QED}}^\mu$ and $J_{\mD}^\mu$.
The equations of motion for the 
field strengths can be written as
\begin{equation}\label{eq:u1EoM}
\matform{K} \partial_\mu \matform{F}^{\mu\nu} = -\matform{J}^{\nu}\ ,
\end{equation}
where
\begin{align}
\matform{A}^\mu := \pmqty{A^\mu \\ A_\mD^\mu},\,\,
\matform{F}^{\mu\nu} := \partial^\mu\matform{A}^\nu - \partial^\nu\matform{A}^\mu,\,\,
\matform{K} := \pmqty{1 & -\epsilon \\ -\epsilon & 1},
\,\,
\matform{J}^\mu := \pmqty{eJ^\mu_{\rqed} \\ e_{\mD}J^\mu_{\mD}}.
\end{align}
Note that $\tM$ does not appear here, as the magnetic mixing is a total derivative in the effective theory.
For a point charge, $\mathcal{J}^\mu(x) = \matform{Q} \delta^\mu_0\delta^3(\mathbf{x})$ where $\matform{Q} = \pqty{e\ncharge, e_{\mD}\ncharge[\mD]}^\top$.%
\footnote{In the dark photon model in Sec.~\ref{sec:U1SU2}, we assume 
no SU(2)$_\mD$ charged fields have U(1)$_\mathrm{QED}$ charge, and hence,
$n_{\mathrm{QED}}^e=0$ or $n_\mD^e = 0$ in the basis of \eqrefer{eq:U1model}. 
However, the interaction energy can be defined for more general cases.}
The static solution in the Coulomb gauge, $\nabla \cdot \vec{\matform{A}} = 0$, is
\begin{equation}
\matform{A}^0 = \frac{1}{4\pi r} \matform{K}^{-1} \matform{Q},\quad\vec{\matform{A}} = 0\ ,
\end{equation}
where $r$ denotes the distance from the point charge.
Therefore, the electric potential energy between two point charges $\matform{Q}$ and $\matform{Q}'$ is given by
\begin{equation} \label{eq:Eint}
E_{\mathrm{int}} 
:=\matform{Q}'^\top \int_r^{\infty}
\dd{x}^i \matform{F}^{0i}
= \matform{Q}'^\top \matform{A}^0 = \frac{1}{4\pi r}\matform{Q}'^\top \matform{K}^{-1} \matform{Q}\ ,
\end{equation}
where $r$ is the distance between the charges.
Here, the electric field is defined by $\matform{E}^i =\matform{F}^{0i} = - \matform{F}_{0i}$. 

To see the effect of the kinetic mixing on the electric potential energy,
let us first consider the case of two QED electric charges.
Plugging in $\matform{Q}=(e\ncharge, 0)^\top$ and $\matform{Q}' = (e\nchargep, 0)^\top$, \eqrefer{eq:Eint} leads to
\begin{equation}
    E_{\mathrm{int}} = \frac{e^2}{1-\epsilon^2}
    \times\frac{\ncharge \nchargep}{4\pi r}\ .
\end{equation}
This is familiar Coulomb's law, except that $e^2$ is replaced with $e^2/(1-\epsilon^2)$.
This deviation is due to the interaction between QED charges via dark photon exchange.

For a dark electric charge and a QED test particle, i.e., $\matform{Q}=(0, e_\mD\ncharge[\mD])^\top$ and $\matform{Q}' = (e\ncharge, 0)^\top$, we have
\begin{equation}
 E_{\mathrm{int}} = \frac{\epsilon ee_\mD }{1-\epsilon^2} \times
 \frac{\ncharge \ncharge[\mD]}{4\pi r} \ .
\end{equation}
Physically, this indicates that 
the QED test charged particle
feels Coulomb force 
from the dark electric charged particle as if it has
 QED electric charge $\epsilon\ncharge[\mD]e_{\mD}/e$.

Note that the definition of the charges depends on the basis of the U(1) gauge fields.
That is, the redefinition $\matform{A}\to\matform{S}\matform{A}$ with a 
 2$\times 2$ regular matrix $\matform{S}$ transforms $\matform{Q}$ to $\matform{S}^{-1\top}\matform{Q}$.
The interaction energy $E_\text{int}$ is, on the other hand, independent of the basis, since it is a physical observable. 
Indeed, the field redefinition also changes $\matform{K}$ 
to $\matform{S}^{-1\top}\matform{K}\matform{S}^{-1}$,
and hence, the interaction energy \eqref{eq:Eint} is intact.

\subsection{Dark Monopoles}
\label{sec:dark monopoles}
Next, we move on to the case with dark magnetic monopoles.
At the phase transition, SU$(2)_\mD\to\text{U}(1)_\mD$, 
the 't Hooft-Polyakov monopole 
can appear~\cite{tHooft:1974kcl,Polyakov:1974ek}. 
In the absence of the kinetic and magnetic mixing terms,
the static configuration
of the monopole at the origin is given by
\begin{align}
\label{eq:heg}
\monohiggs^a=v_1 H(r)\displaystyle{\frac{x^a}{r}}\ , \quad
A_{\mD 0}^a = 0 \ ,\quad
A^{a}_{\mD i}=
\frac{1}{e_\mD}\displaystyle{\frac{\epsilon^{aij}x^j}{r^2}}F(r)\ ,\quad (i,j=1,2,3)\ ,
\end{align}
where $r=\sqrt{x^2+y^2+z^2}$.
The profile functions $H(r)$ and $F(r)$ satisfy the boundary conditions
\begin{alignat}{2}
H(r)&\to \text{const.}\times r\ ,\, (r\to0)\ , &\qquad H(r)&\to 1\ ,\, (r\to \infty)\ ,\\
F(r)&\to \text{const.}\times r^2 \ , \, (r\to 0)\ , & \qquad F(r)&\to 1\ ,\, (r\to \infty)\ ,
\end{alignat}
where they approach their asymptotic values exponentially at $r\to \infty$.

To see the magnetic field,
it is convenient to define the effective U(1)$_\mD$
field strength as
\begin{equation}
\label{eq:effectiveF}
F_{\mD\mu\nu} := \frac{1}{v_1}\monohiggs^a F^{a}_{\mD\mu\nu}
\end{equation}
(see e.g. Ref.~\cite{Shifman:2012zz}).
The only non-vanishing components of $F_\mD^{\mu\nu}$ are 
\begin{equation}
    F_\mD^{i j } = -
    \frac{1}{e_\mD}
\frac{\epsilon^{ijk}x^k}{r^3} (2F-F^2)H \ , \quad (i,j=1,2,3)\ .
\end{equation}
Hence, the dark magnetic charge of the monopole solution is given by
\begin{equation}
\label{eq:Mcharge}
    Q_{\mD}^m:=  \int_{r \to \infty} 
    \dd[2]
S_{i} B_\mD^{i  } =  -\frac{4\pi}{e_\mD} \ ,
\end{equation}
where $\dd[2]{S}_{i}$ is the surface element of a two dimensional sphere surrounding the monopole.

Now, let us consider the effect of the kinetic and magnetic mixings.
As we assume those parameters to be tiny, their effects on the configuration \eqref{eq:heg} can be safely neglected.
\modifiedat{PRDv2}{(For the stability of the topological defects in the presence of the mixing terms, see the Appendix~\ref{sec:stability}.)}
The equation of motion for $A_\mu$ in the $\uone\times\suD$ theory is
\begin{equation}\label{eq:U1EoM}
    \partial_{\mu} F^{\mu\nu}
     -\epsilon \partial_{\mu} 
      F_{\mD}^{\mu\nu}
      + \frac{\tM}{8\pi^2} \partial_\mu \tilde{F}_\mD^{\mu\nu}
      =0 \ .
\end{equation}
The third term vanishes at $r\gg (e_\mD v_1)^{-1}$ due to the 
Bianchi identity of the effective U(1)$_\mD$ theory.
In the vicinity of the monopole $r\sim \order{(e_\mD v_1)^{-1}}$, on the other hand, it does not vanish where
\begin{equation}
\label{eq:QEDandMonopole}
\partial_{\mu}\tilde{F}_{\mD}^{\mu\nu}
= \frac{1}{v_1} \partial_{\mu}
(\monohiggs^a \tilde{F}^{a\mu\nu}_{\mD})
= \frac{1}{v_1} (D_{\mu}
\monohiggs)^a \tilde{F}^{a\mu\nu}_{\mD}\neq 0\ .
\end{equation}
In the last equality, we used the Bianchi identity of SU$(2)_\mD$, i.e., $D_\mu\tilde{F}_{\mD}^{a\mu\nu}=0$.
Besides, the effective field strength $F_{\mD}^{\mu\nu}$ satisfies $\partial_\mu F_\mD^{\mu\nu}=0$ even at $r\to 0$, and hence, the second term 
in \eqrefer{eq:U1EoM} vanishes.

As a result, the equation of motion for the QED electric field is given by
\begin{equation}
\label{eq:Gauss}
\partial_i E^i =\frac{\tM}{8\pi^2} \partial_i B_{\mD}^i.
\end{equation}
Thus, we find the solution of Eq.\,\eqref{eq:U1EoM} in the Coulomb gauge to be
\begin{equation}
\label{eq:monopole A0}
     A^0 \simeq- \frac{\tM Q_\mD^m}{8\pi^2} \times \frac{1}{4\pi r} =
     \frac{\theta_\mathrm{mix}}{8\pi^2 e_\mD}\times \frac{1}{r}
     \ ,\quad 
    \vec{A}=0 \ ,
\end{equation}
for $r \gg (e_\mD v_1)^{-1}$
to the leading order of the mixing parameters.

Accordingly, the interaction energy between
a QED test particle with $\mathcal{Q}=(\ncharge,0)^\top$ and a dark monopole is given by,
\begin{equation}
    \label{eq:tMEint}
    E_{\mathrm{int}} = - \frac{e\tM Q_\mD^m \ncharge}{8\pi^2} \times \frac{1}{4\pi r}    
     \ ,
\end{equation}
for $r \gg (e_\mD v_1)^{-1}$ to the leading order of the mixing parameters.
This shows that the dark magnetic monopole exerts Coulomb force to QED particles through the magnetic mixing, whereas the kinetic mixing induces no interactions between them~\cite{Brummer:2009oul}.

\subsection{Dark Dyons}
The $\suD$ sector admits dyons, magnetic monopoles that also have electric charge~\cite{Julia:1975ff}.
The dyon solution is described by \eqrefer{eq:heg} but with $A_{\mD 0}^a$ replaced by
\begin{equation}
    \label{eq:dyon}
    A^a_{\mD 0} = \frac{1}{e_\mD}\frac{x^a}{r^2}J(r)\ .
\end{equation}
The boundary conditions for $J(r)$ are
\begin{equation}
        J(r) \to \text{const.}\times r^2\ , \,  (r\to 0) \ , \qquad
        J(r) \to Mr + b\ ,  \, (r\to\infty)\ ,
\end{equation}
where $M$ and $b$ are the  
parameters with mass dimensions one and zero, respectively.

The dark magnetic field $F_{\mD ij}$ is not modified by \eqrefer{eq:dyon}.
On the other hand, the dark electric field no longer vanishes:
\begin{equation}
    F_{\mD}^{0i} = \frac{1}{e_\mD}\frac{x^i}{r}\dv{r} \frac{J(r)}{r} \to -\frac{b}{e_\mD}\frac{x^i}{r^3}\ .
\end{equation}
Hence, the dark electric charge of the dyon is found to be
\begin{equation}
    \label{eq:dyon_echarge}
    \qcharge[\mD] = -\frac{4\pi b}{e_\mD} = bQ^m_\mD
\end{equation}
in the absence of the mixing to the QED sector.

By remembering how the dark electric charges and dark magnetic charges induce the Coulomb force on QED charged particles (see Eq.\,\eqref{eq:U1EoM}), we find 
the interaction energy to be
\begin{equation}
    \label{eq:Eint_dyon}
    E_{\mathrm{int}} = e\ncharge\pqty{-\frac{\tM}{8\pi^2}Q_\mD^m + \epsilon \qcharge[\mD]}\times \frac{1}{4\pi r}\ ,
\end{equation}
to the leading order in the mixing parameters.

\begin{figure}
    \centering
    \includegraphics[width=\linewidth]{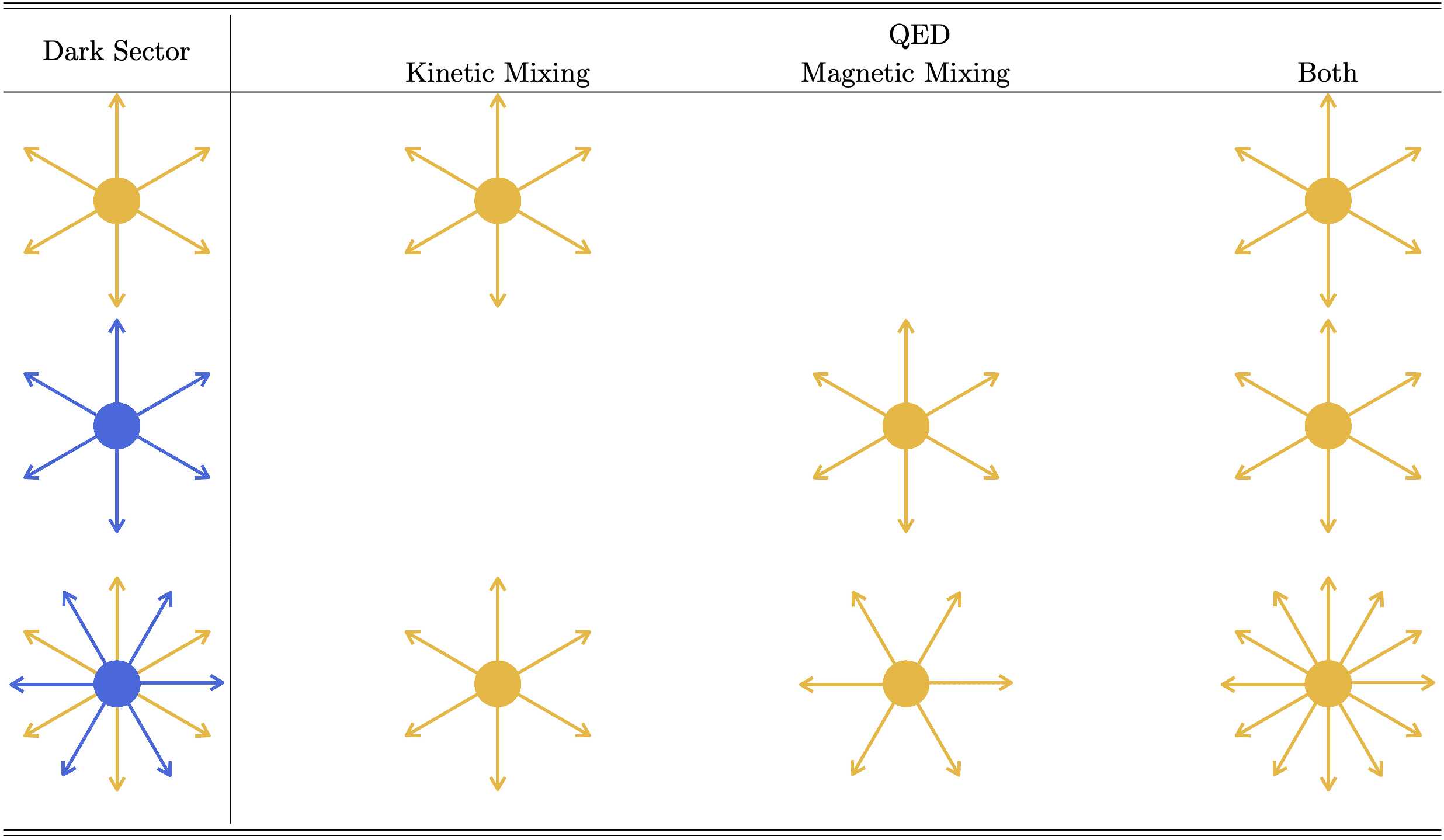}
    \caption{Summary of the appearance of dark charges as QED charges. The yellow lines indicate dark/QED electric field and blue lines indicate dark magnetic field. The leftmost column shows objects in the dark sector and the other columns describe QED electric fields induced by the mixing terms.}
    \label{tab:U1symmetric}
\end{figure}

This concludes our analysis on the interactions between dark charge objects and QED charges in the $\uone[\mD]$ symmetric phase.
Fig.~\ref{tab:U1symmetric} summarizes the results in this section.

\subsection{Charge Quantization}
The dark magnetic charge is quantized as it corresponds to the 
topological number $n_\mD^m \in \mathbb{Z}$ of the configuration, with which $Q^m_\mD = -4\pi n_\mD^m/e_\mD$.
Its quantization is not affected by the mixings to the QED sector.

The dark electric charge is arbitrary at the classical level, as in \eqrefer{eq:dyon_echarge}.
In a quantum theory, however, the dyon electric charge has to be quantized~\cite{Tomboulis:1975qt,Gervais:1976ec}.
To see this in our setup, let us consider 
the residual global $\uone[\mD]$ symmetry around $\monohiggs$,
\begin{align}
\label{eq:U1D trans}
    \delta A^a_{\mD\mu} = -\frac{1}{e_\mD v_1}D_\mu \monohiggs^a\ , \quad
    \delta A_\mu = 0 \ ,\quad
    \delta \monohiggs^a = 0\ .
\end{align}
As shown in  Appendix~\ref{sec:noether}, the corresponding Noether charge is given by
\begin{equation}
    N_{\text{U(1)}_\mD}= \frac{1}{e_\mD}\qcharge[\mD] - \frac{\epsilon}{e_\mD}\qcharge - \frac{\theta_\mD e_\mD}{8\pi^2}Q^m_\mD\ .
\end{equation}
The electric and magnetic charges are measured by electric flux,  
\begin{align}
     \pmqty{\qcharge \\ \qcharge[\mD]} := \int \dd[2]{S}_i \matform{E}^i \ ,
\end{align}
and the magnetic flux (see Eq.\,\eqref{eq:Mcharge}).
Since $N_{\uone[\mD]}$ is one of the generators 
of global SO(3)$_\mD\simeq\,\,$SU(2)$_\mD$ transformation,
we find $ N_{\text{U(1)}_\mD} \in \mathbb{Z}$,
which constrains $\qcharge[\mD]$ of dyons~\cite{Witten:1979ey,Salam:1980ri} (see also Ref.~\cite{Balachandran:2018cgw}).
Note that this is the usual Witten effect in the absence of the mixings.

Let us also comment on the effects of the $\theta$-terms $\mathcal{L}_{\theta}$ to the equations of motion.
In our formulation, the $\uone\times$SU(2)$_\mD$ gauge potentials $A_\mu$ and $A_{\mD\mu}^a$ are globally defined, and hence,  $\mathcal{L}_{\theta}$ does not affect the equations of motion.
In the U$(1)_\mathrm{QED}\times$U$(1)_\mD$ formulation,
on the other hand, it is 
also possible to introduce monopoles 
as a singularity~\cite{Brummer:2009oul}.
In this treatment, $\mathcal{L}_{\theta}$
classically induces an electric field around a dark monopole
(see also Ref.~\cite{Coleman1983}).%
\footnote{
Strictly speaking, 
singularities in the dark sector obscure the boundary condition of the QED gauge potential.
Our treatment based on the $\uone\times\suD$ theory does not have such subtleties.}

\section{\texorpdfstring{$\mathrm{U(1)}_\mD$}{U(1)D} Broken Phase}
\label{sec:U1broken}
\subsection{Dark Elementary Charged Particles}
Let us consider the case without monopoles, where the effective theory \eqref{eq:U1model} is valid.
At the trivial vacuum \eqref{eq:U1VEV},
the $\uone[\mD]\times\uone$
model is reduced to
\begin{align}
\label{eq:dark photon}
\mathcal{L}=&-\frac{1}{4}F_{\mu\nu}F^{\mu\nu}-\frac{1}{4}{F_\mD}_{\mu\nu}F_\mD^{\mu\nu}
- \frac{1}{2}m_{\mD}^2 A_{\mD\mu}A^{\mu}_{\mD}
+\frac{\epsilon}{2}F_{\mu\nu}F_\mD^{\mu\nu} \cr
&
-\frac{\tM}{16\pi^2}F_{\mu\nu}\tilde{F}_{\mD}{}^{\mu\nu}
- \frac{e^2\theta}{32\pi^2}F_{\mu\nu}
\tilde{F}^{\mu\nu} 
- \frac{e_\mD^2\theta_\mD}{32\pi^2}F_{\mD\mu\nu}
\tilde{F}_\mD^{\mu\nu}
\cr
&+e A_\mu J_\mathrm{QED}^\mu 
+ e_{\mD} A_{\mD\mu} J_\mD^\mu
\ ,
\end{align}
where $m_\mD^2 = 2 e_{\mD}^2 v^2$.

In this case, it is most convenient 
to introduce a new basis
\begin{align}
\label{eq:decoupled}
 \pmqty{A^\mu \\ A_\mD^\mu}=:\pmqty{1 & \frac{\epsilon}{\sqrt{1-\epsilon^2}} \\ 0 & \frac{1}{\sqrt{1-\epsilon^2}} } \pmqty{A^{\prime\mu} \\ A_\mD^{\prime\mu}}\ , 
\end{align}
with which the equations of motion are given by
\begin{align}
    \label{eq:broken EoM}
    &\partial_\mu F^{\prime\mu\nu}  = e J_\mathrm{QED}^\nu\ , \\
    &\partial_\mu F_\mD^{\prime\mu\nu} - m_\mD^{\prime 2} 
    A_\mD^{\prime \nu} = \frac{e_\mD}{\sqrt{1-\epsilon^2}}J_\mD^\nu
    + \frac{\epsilon e}{\sqrt{1-\epsilon^2}}J_\mathrm{QED}^\nu
    \ ,
\end{align}
where $m_\mD^{\prime 2} = m_{\mD}^2/(1-\epsilon^2)$.
We refer to the bases $(A_\mu, A_{\mD, \mu})$ and $(A'_\mu, A_{\mD\mu}')$ the original and decoupled bases, respectively.

Then the interaction energy
between a dark electric charge and a QED test particle, i.e., $\matform{Q}=(0, e_\mD\ncharge[\mD])^\top$ and $\matform{Q}' = (e\ncharge, 0)^\top$ 
in the original basis,
is suppressed by $e^{-m_\mD'r}$:
\begin{equation}
\label{eq:charge in broken U1}
 E_{\mathrm{int}} = \frac{\epsilon ee_\mD }{1-\epsilon^2} \times
 \frac{\ncharge \ncharge[\mD]}{4\pi r}e^{-m_\mD'r} \ .
\end{equation}
Note that the $\theta$-terms in \eqrefer{eq:U1model} has no observable effect in this case.

\subsection{Dark Strings}
\label{sec:dark strings}
Let us continue to assume the absence of monopoles.
However, we now consider the vacuum configuration of $\uone[\mD]$ breaking associated with a string 
as discussed in Ref.~\cite{Hiramatsu:2021kvu}.
We continue to use the decoupled basis.
The static string solution along the $z$-axis is given by the form (see e.g., Ref.~\cite{Vilenkin:2000jqa})
\begin{align}
\label{eq:string ansatz1}
\uphi &=v h(\rho)e^{i n\azimuth}\ , \\
\label{eq:string ansatz2}
A'_{\mD i}&=-\frac{n}{e_\mD'}\frac{\epsilon_{ij}x^j}{\rho^2}f(\rho)\ ,~~~~(i,j=1,2)\ , \\
A'_{\mD 0} &= A'_{\mD 3} = 0 \ ,
\end{align}
where $n \in \mathbb{Z}$ is the winding number of the string configuration, $h(\rho)$, $f(\rho)$ the profile functions,
and $e'_\mD= e_\mD/\sqrt{1-\epsilon^2}$.
The cylindrical coordinate is given by $\azimuth={\arctan}(y/x)$ and $\rho=\sqrt{x^2+y^2}$.
The two-dimensional anti-symmetric tensor is defined by $\epsilon_{12}=1$.%
    \footnote{Noting that $\dd{\azimuth} = - \dd{x}^i \epsilon_{ij}x^j/\rho^2$, 
Eq.\,\eqref{eq:string ansatz2} can be rewritten by $A'_{\mD i} \dd{x}^i = n/e_\mD' \times f(\rho)\dd{\azimuth}$.} 
The boundary conditions for the profile functions are
\begin{align}
h(\rho)\rightarrow 0\ , ~(\rho\rightarrow0)&\ ,~~~~~h(\rho)\rightarrow 1\ , ~(\rho\rightarrow \infty)\ ,\\
f(\rho)\rightarrow 0\ , ~(\rho\rightarrow 0)&\ ,~~~~~f(\rho)\rightarrow 1\ , 
~(\rho\rightarrow \infty)\ .
\end{align}
They approach unity for $\rho \gg (e'_\mD v)^{-1}$ exponentially.
The winding number is related to the dark magnetic flux along the string core by
\begin{align}
\label{eq:Wilson0}
\int \dd[2]{x}B_{\mD3}'=\oint_{\rho\rightarrow \infty} A'_{\mD i} \dd{x}^i=\frac{2\pi n}{e_\mD'}\ .
\end{align}

In the decoupled basis, the absence of the kinetic mixing implies $A'_\mu = 0$.
Nevertheless, QED test charges 
defined 
in the original basis
feel the Aharonov-Bohm (AB) effect
through $A_{\mu}\neq 0$. The corresponding AB phase around the string is given by~\cite{Hiramatsu:2021kvu}
\begin{align}
\label{eq:ABphase}
\ncharge W_{\mathrm{QED}} = \frac{\ncharge\epsilon e}{\sqrt{1-\epsilon^2}}\oint A'_{\mD\mu} \dd{x}^\mu= \frac{2\pi n \ncharge q \epsilon e}{e_\mD} \ .
\end{align}

As in the case of elementary dark charges, 
the $\theta$-terms do not affect the equations of motion because of the $\uone$ and $\uone[\mD]$ Bianchi identities.
Thus, they do not modify the field configurations, and hence, the AB phases.

It is also instructive to see the dark string 
in the original basis.
Substituting \eqrefer{eq:string ansatz2} into \eqrefer{eq:decoupled}, we find
\begin{align}
A_{i}&=-\frac{\epsilon n}{e_\mD}\frac{\epsilon_{ij}x^j}{\rho^2}f(\rho)\ , \\
A_{\mD i}&=-\frac{n}{e_\mD}\frac{\epsilon_{ij}x^j}{\rho^2}f(\rho)
\end{align}
for $i,j=1,2$.
In this picture, $A_i$ is induced by 
the $\uone[\mD]$ current of $\uphi$,
\begin{align}
    J_\mathrm{\uphi}^i= i \uphi D^i\uphi^\dagger - i \uphi^\dagger D^{\modifiedat{PRDv4}{i}} \uphi = 2v^2 n \frac{\epsilon^{ij}x^j}{\rho^2}h^2(f-1)\ ,
\end{align}
through the kinetic mixing.
This expression allows us to interpret the AB effect on QED charges as a result of a solenoid around the string.

\subsection{Dark Beads}
\label{sec:dark beads}
\subsubsection{Dark beads configuration}
In this section, we consider 
the effects of the so-called bead solution which appears in the $\uone[\mD]$ broken phase around a dark magnetic monopole without electric charge.%
\footnote{This assumption requires $\theta_\mD = 0$.}
Here, we begin with a review of the bead solution without mixing to the QED sector (see Ref.~\cite{Kibble:2015twa} for a review).

As we have seen in Sec.~\ref{sec:U1SU2}, 
$\strhiggs$ prefers to be orthogonal to $\monohiggs$ because of the $\kappa$ term in the potential \eqref{eq:potential}.
However, such a configuration of $\strhiggs$ with 
a constant amplitude, $|\strhiggs| = v_2$,
is impossible due to the
Poincar\'e--Hopf (hairy ball) theorem
around the monopole solution \eqref{eq:heg}.
Rather, $|\strhiggs|$ should vanish at some points at $r\to \infty$ and strings must extend in those directions.
Such a configuration is called a beads solution~\cite{Hindmarsh:1985xc,Everett:1986eh,Aryal:1987sn,Kibble:2015twa}.
A network of connected bead solutions is also called a necklace~\cite{Berezinsky:1997td}.%
\footnote{Necklace solutions in SO$(10)$ and E$_6$ are discussed in e.g. Ref.~\cite{Lazarides:2019xai}.}

To see the formation of beads, it is helpful to consider a monopole in a gauge defined in two slightly overlapping charts covering the northern and southern hemispheres,
\begin{align}
\label{eq:NSchart}
    U_{N} &= 
    \left\{(r, \zenith,\azimuth)|0 \le \zenith \le {\pi}/{2} + \varepsilon
    ,\, r>R\right\}\\
    U_{S} &= 
    \left\{(r, \zenith,\azimuth)|{\pi}/{2}-\varepsilon \le \zenith \le \pi
    ,\, r>R\right\}\ .
\end{align}
Here, $\zenith$ is the 
zenith angle, $\varepsilon$ is a small positive parameter, and $R\gtrsim (e_\mD v_1)^{-1}$.
In each chart, we transform the monopole solution \eqref{eq:heg} by
\begin{align}
\label{eq:combedgauge}
    &g_N =\left(
    \begin{array}{cc}
      c_{\zenith/2}  &  e^{-i\azimuth}s_{\zenith/2} \\
      - e^{i\azimuth}s_{\zenith/2}& c_{\zenith/2}
    \end{array}
    \right)\ ,
    ~~~~~~~ g_S =\left(
    \begin{array}{cc}
       e^{i\azimuth}c_{\zenith/2}  & s_{\zenith/2} \\
      -s_{\zenith/2}& e^{-i\azimuth}c_{\zenith/2}
    \end{array}
    \right)\ ,
\end{align}
that is,%
\begin{align}
    \monohiggs^a t^a &\to  \monohiggs{}_{N,S}^a t^a = g_{N,S} \monohiggs^a t^a g_{N,S}^\dagger\ , \\
A^a_{\mD i}t^a &\to  A_{\mD N,S\,i}^a t^a = g_{N,S} A_{\mD i}^at^a g_{N,S}^\dagger +
\frac{i}{e_\mD} g_{N,S} \partial_ig_{N,S}^\dagger\ ,
\label{eq:localDP}
\end{align}
with $t^a$ ($a=1,2,3$) being the halves of the Pauli matrices.
We call this gauge choice the combed gauge.

In this gauge, the asymptotic behavior of the monopole at $r \gg (e_\mD v_1)^{-1}$ is given by%
\footnote{Here, we denote the gauge potentials as one-form gauge fields.}
\begin{align}
\label{eq:phiN}
    \monohiggs{}_N^a &\to v_1 \delta^{a3}\ ,\\
    A^{a}_{\mD N} &\to \frac{1}{e_\mD}\delta^{a3} (\cos\zenith -1) \dd{\azimuth}
\end{align}
in the $U_N$ chart and
\begin{align}
\label{eq:phiS}
    \monohiggs{}^a_S  &\to v_1 \delta^{a3}\ ,\\
    A^{a}_{\mD S}&\to \frac{1}{e_\mD}\delta^{a3} 
    (\cos\zenith+1)  \dd{\azimuth}\ ,
\end{align}
in the $U_S$ chart, while
 \modifiedat{PRDv4}{$A^{a}_{\mD N,S}$} vanish asymptotically.

 In the combed gauge, $A^3_{\mD N,S}$ in each chart are connected with each other at around the equator $\zenith\sim \pi/2$  by
\begin{align}
    A^{3}_{\mD S}= A^{3}_{\mD N}+ \frac{2}{e_\mD} \dd{\azimuth}\ .
\end{align}
That is, the gauge transition function connecting the two charts is
\begin{align}
    \label{eq:transition}
    t_{NS} = e^{2i\azimuth}\ .
\end{align}

\begin{figure}[tbp]
 \begin{minipage}{0.45\linewidth}
\includegraphics[width=0.3\linewidth]{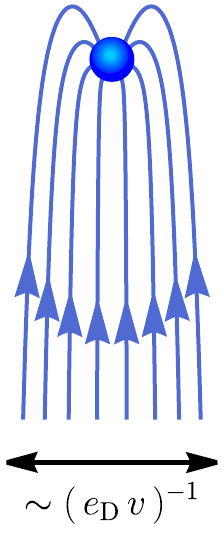}
 \end{minipage}
 \begin{minipage}{0.45\linewidth}
\includegraphics[width=0.23\linewidth]{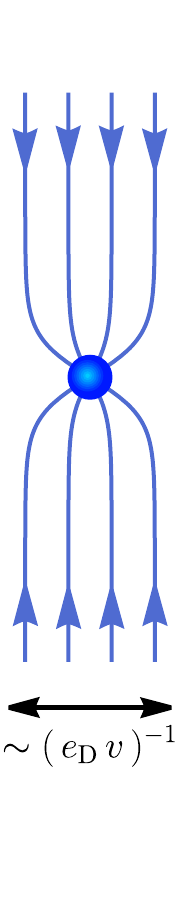}
\end{minipage}
\caption{Schematic pictures of 
the bead solutions.
The ball denotes the dark magnetic monopole, and the arrows denote the dark magnetic field. 
Left) 
The attached string with $n=2$
extends in the negative $z$ direction.
Right) 
The attached string with $n=-1$
extends in the positive $z$ direction while the one with $n=1$
extends in the negative $z$ direction.
}
\label{fig:beads}
\end{figure}

Now we discuss the winding of $\uphi$.
First, let us suppose that ${\uphi}$ takes a constant expectation value $v$
in the northern hemisphere for $r \gg (e_\mD v_1)^{-1}$.
Then the U(1)$_\mD$ magnetic flux is expelled from the northern hemisphere by the Meissner effect, and hence,
the gauge potential in the northern hemisphere is trivial:
\begin{equation}
    A^{3}_{\mD Ni} = 0
\end{equation}
for $r\gg(e_\mD v_1)^{-1}$.
In the overlapping region, the scalar and gauge fields in the $U_S$ chart take the form
\begin{align}
\label{eq:phiS2}
    {\uphi}_S &= e^{2i\azimuth} \uphi_N \ , \\
     A_{\mD S}^3 &= A_{\mD N}^3 + \frac{2}{e_\mD}\dd \azimuth
\end{align}
for $r\gg (e_\mD v)^{-1}$ due to the non-trivial transition function \eqref{eq:transition}.
This shows that the trivial configuration in the northern hemisphere requires a non-trivial winding of $\uphi_S$.
Note that the minimum energy solution of U(1)$_\mD$ with a non-trivial winding is a string 
with a radius of $\order{(e_\mD v)^{-1}}$.
Thus, \eqrefer{eq:phiS2}
shows that a string with $n=2$ is formed in the southern hemisphere.
The dark magnetic flux for the $n=2$ string is
\begin{align}
\label{eq:fluxS}
    \oint  A^{3}_{\mD Si}\dd{x}^i = \frac{4\pi}{e_\mD}\ ,
\end{align}
which coincides with the total magnetic flux of the monopole.
As a result, we find that the magnetic flux of the magnetic monopole escapes through the string (see the left panel of Fig.\,\ref{fig:beads}).
This configuration is consistent with the Poincar\'e--Hopf theorem since $\strhiggs^a = 0$ at the center of the string.%
\footnote{This configuration is not static, and the dark monopole is pulled in the negative $z$ direction.}

Next, let us 
consider an $n = -1$ string in the northern hemisphere extending from the monopole toward $z\to+\infty$. 
The asymptotic behavior of the string for $\rho\gg (e_\mD v)^{-1}$ is given by
\begin{align}
\label{eq:phiNstring}
    \uphi_N &\to ve^{-i\azimuth}  \ , \\
    A_{\mD N}&\to -\frac{1}{e_\mD}\dd{\azimuth}\ .
\end{align}
The corresponding asymptotic behavior in the southern hemisphere is
\begin{align}
\label{eq:phiSstring}
      \uphi_{S} &= e^{2i\azimuth} \uphi_N \to ve^{i\azimuth}  \ , \\
    A_{\mD S}&=A_{\mD N } + \frac{1}{e_\mD}\dd{\azimuth}\  \to \frac{1}{e_\mD}\dd{\azimuth}\ ,
\end{align}
namely the string solution with $n=1$.
Thus, in this configuration, 
a string and an anti-string are attached to a magnetic monopole
 (see the right panel of Fig.\,\ref{fig:beads}).
The magnetic flux confined in the string and the anti-string is given by
\begin{align}
\label{eq:beadflux}
-  \oint A^{3}_{\mD Ni}\dd{x}^i +\oint A^{3}_{\mD Si}\dd{x}^i =   \frac{4\pi}{e_\mD} \ ,
\end{align}
which coincides with the magnetic flux of the monopole.
This configuration is called the bead solution~\cite{Hindmarsh:1985xc}.

\subsubsection{Kinetic mixing}
So far in this subsection, we have ignored the mixing terms.
As discussed in Ref.\,\cite{Hiramatsu:2021kvu}, the 
kinetic mixing induces a non-trivial QED magnetic field called pseudo-monopoles.

As we saw in Sec.~\ref{sec:dark strings}, the strings attached to the monopole induces QED magnetic field along them.
Thus, we find that the QED magnetic flux (in the original basis)
flows into the magnetic monopole:
\begin{equation}
\label{eq:QEDbeadflux}
  \left[-\oint A^{3}_{Ni}\dd{x}^i +\oint A^{3}_{Si}\dd{x}^i\right]_{\text{string}} =   \frac{\epsilon 4\pi}{e_\mD} \ .
\end{equation}

\begin{figure}[tbp]
\includegraphics[width=0.23\linewidth]{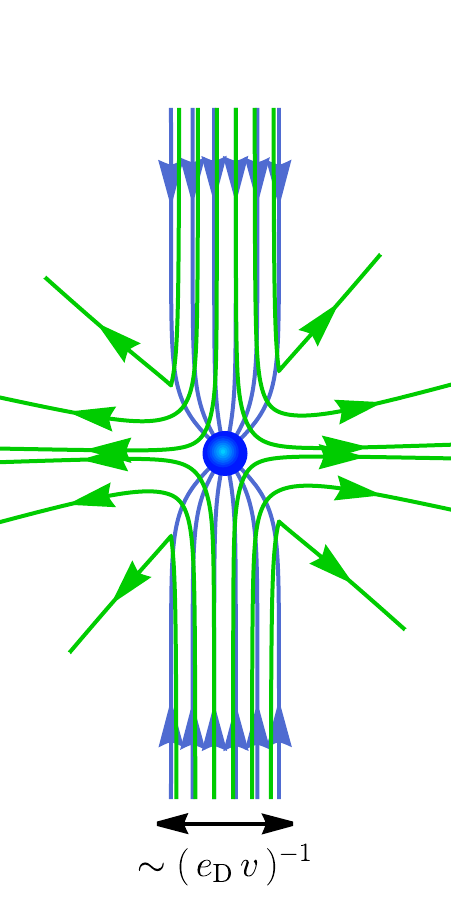}
\caption{Schematic picture of a QED pseudo-monopole.
The ball denotes a dark magnetic monopole, and the blue arrows denote the dark magnetic field. 
The green arrows denote the QED magenetic field of the pseudo-monopole, which satisfies the Bianchi identity.
}
\label{fig:pseudomonopole}
\end{figure}

In the original basis, however, the QED Bianchi identity prohibits sources and sinks of the QED magnetic field.
Since the QED magnetic flux \eqref{eq:QEDbeadflux} is confined within the strings at $|z|\gg (e_\mD v_1)^{-1}$, the incoming flux \eqrefer{eq:QEDbeadflux} must leak at the ends i.e., in the vicinity of the monopole:
\begin{equation}
\int \dd[2]{S}^{i}B_{i}\big|_{\text{leak}} = \frac{\epsilon 4\pi}{e_\mD}\ .
\end{equation}
Since the leakage occurs
from the tiny region $r=\order{(e_\mD v_1)^{-1}}$, the magnetic flux should be spherical for large $r$, and hence,
\begin{align}
    B_i\big|_{\text{leak}} = 
     \frac{\epsilon}{e_\mD}\frac{ x_i}{r^3}\ ,
\end{align}
which looks like a QED monopole (see Fig.~\ref{fig:pseudomonopole}).
We call this pseudo-monopole.

So far, no analytic expressions for the bead nor the pseudo-monopole have been known.
However, their formation is confirmed by classical lattice simulation \cite{Hindmarsh:2016dha, Hiramatsu:2021kvu}.

\subsubsection{Magnetic mixing}
Next, let us discuss the effect
of the magnetic mixing $\tM$ while we set $\epsilon = 0$. 
In this case, the equation of motion for $A_\mu$ is given by
\begin{equation}\label{eq:U1EoM2}
\partial_i E^i =\frac{\tM}{8\pi^2} \times  \partial_i B_{\mD}^i\ ,
\end{equation}
(see Eq.\,\eqref{eq:U1EoM}).
Since the contribution of $\tM$ is proportional to $\partial_i B_{\mD}^i$, only dark monopoles contribute to the QED electric field even in the case of the dark bead solution.
Therefore, we have again the QED electric potential \eqref{eq:monopole A0} and the interaction energy \eqref{eq:tMEint}.
Notice that this interaction energy 
is not suppressed by $e^{-m_{\mD}'r}$
even in the U(1)$_\mD$ broken phase, unlike the case of dark elementary charges (see \eqrefer{eq:charge in broken U1}). 
As a result, we find that the dark magnetic mixing induces a spherical Coulomb potential around the dark monopole even though the dark magnetic flux is confined into the strings.

When both the kinetic and magnetic mixing exist, 
the dark bead configuration induces the QED pseudo-monopole and spherical QED Coulomb force simultaneously at the leading order of the mixing parameters.

\subsection{Dark Dyonic Beads}
\subsubsection{Dark dyonic beads configuration}
In this section, we qualitatively describe the case where the original dark monopole also has dark electric charge.
The dark magnetic flux of the dyon demands the formation of the bead solution in the $\uone[\mD]$ broken phase,
as in the case of dark monopoles.

The dark electric field, on the other hand,  decays as $\sim e^{-m_\mD' r}$ due to the 
the mass term in \eqrefer{eq:broken EoM}.
Note however that since $\uone[\mD]$ is restored at the string core, the dark electric field is no longer spherical and takes a rugby ball-like configuration along the 
dark strings.
For detailed structure of the solution, 
we need numerical simulation 
which will be discussed elsewhere.

\subsubsection{Interactions through the mixing terms}
Finally, let us discuss the effects of the 
mixing terms.
To the linear order of the mixing parameters,
the effects of the dark dyonic beads can be described by the superposition of those of dark beads and a dark electric charge.

As we have seen in the previous section,
the beads part induces a pseudo-monopole 
through the kinetic mixing and induces a 
QED Coulomb potential through the magnetic mixing.
On the other hand,
the electric charge part
induces a non-spherical decaying potential
for QED charges through the kinetic mixing, while the magnetic mixing does nothing.
The resultant interaction energy is given by
\begin{align}
E_{\mathrm{int}} = 
\left(- \frac{\tM}{8\pi^2} Q_\mD^m  + 
 \epsilon  \qcharge[\mD] e^{-\tilde{m}_\mD'r}\right)\times \frac{e\ncharge}{4\pi r}
     \ ,
\end{align}
where $r$ and $\zenith$ dependent mass $\tilde{m}_\mD'$ accounts for the distortion of the decaying potential.

\begin{figure}[tbp]
    \centering
    \includegraphics[width=\linewidth]{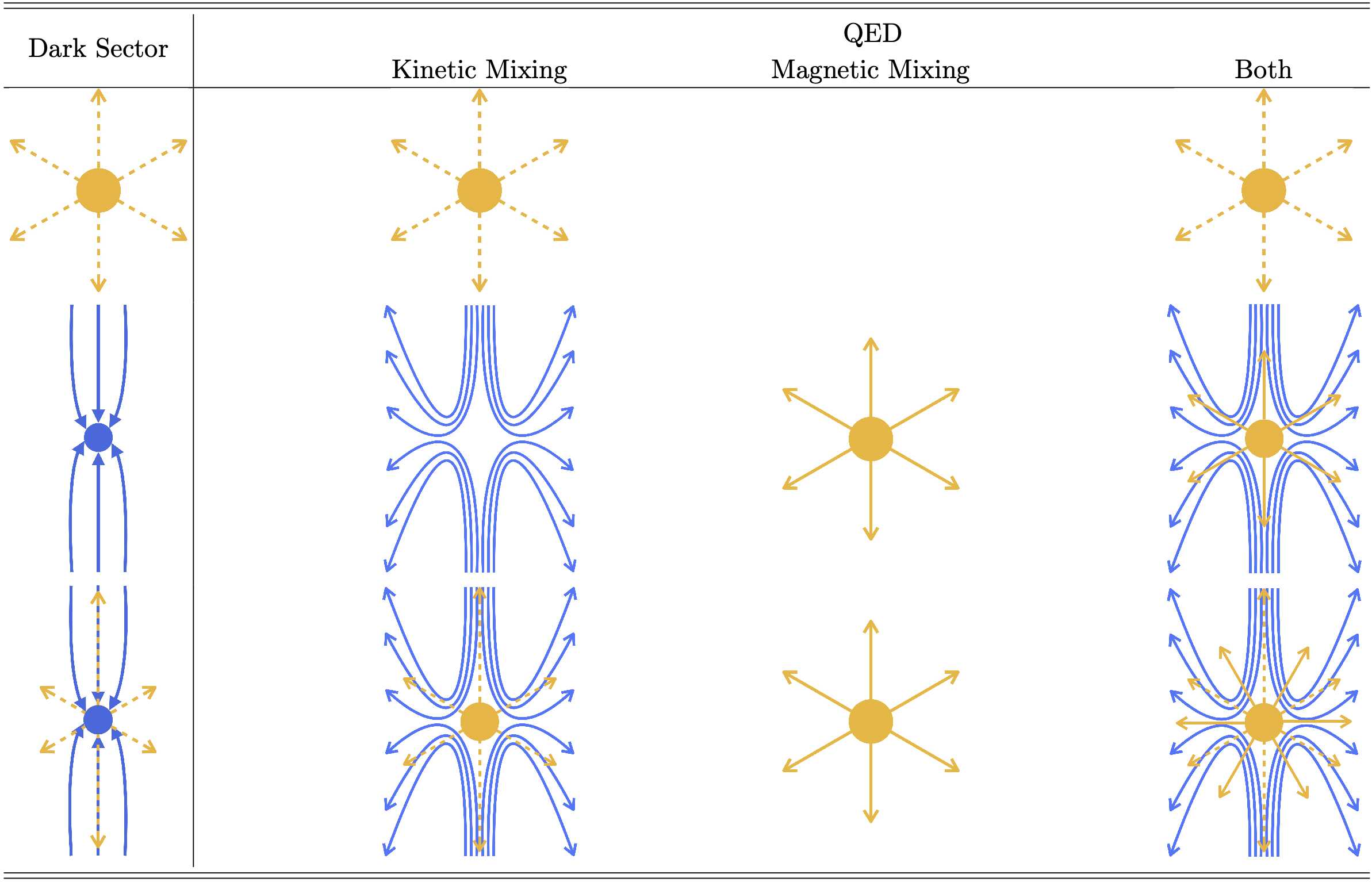}
    \caption{Summary of the appearance of dark objects as QED objects in the $\uone[\mD]$ broken phase. The yellow lines indicate dark/QED electric field and blue lines indicate dark/QED magnetic field. The leftmost column shows objects in the dark sector and the other columns describe QED electromagnetic fields induced by the mixing terms. Dashed lines indicates exponential decay of the field.
    \protect\modifiedat{PRDv2}{
    Notice that the QED electric field induced by the dark elementary charged particles is absent in the decoupled basis in \eqrefer{eq:decoupled}.}} 
    \label{tab:U1broken}
\end{figure}

This concludes our analysis on $\uone[\mD]$ broken phase.
We show the summary of the QED field strengths that the QED charged particle feel in Fig.~\ref{tab:U1broken}.

\subsection{QED Electric Charge Conservation}
One may wonder whether the QED electric charge is conserved when a dark monopole forms.
To clarify this point, two definitions of the electric charge must be carefully distinguished: $\ncharge$, the $\uone$ quantum number, and $\qcharge$, the charge measured by the field strength.

$\ncharge$ is conserved by Noether's theorem.
By definition, monopoles have no contribution (see also Appendix~\ref{sec:noether}).

On the other hand, \eqrefer{eq:Gauss} shows that $\qcharge$ induced by magnetic mixing is proportional to $Q^m_\mD$ even in the $\suD$ symmetric phase. 
The magnetic charge has an associated current conserved throughout the evolution:
\begin{align}\label{eq:magcurrent}
    Q^m_{\mD} &= \int \dd[3]{x} J_{\mathrm{M}, \mD}^{\modifiedat{PRDv3}{0}} \\
    J_{\mathrm{M}, \mD}^\mu&:=-\frac{1}{2}\epsilon^{\mu\nu\rho\sigma}\partial_\nu \pqty{\frac{\monohiggs^a}{v_1}F^a_{\mD\,\rho\sigma}}\ .
\end{align}
Thus, monopole formation does not create any extra QED electric charge.
Rather, the monopole electric charge is just a concentration of already existing charge.

\section{Conclusions}
In this paper, we studied the effects of the dark objects on the QED sector through the mixing between the dark photon and the QED photon, where the dark photon 
appears as a result of the successive symmetry breaking $\suD \to\uone[\mD]\to \mathbb{Z}_2$.
We extended the previous analysis in Ref.~\cite{Hiramatsu:2021kvu} by newly 
considering the effects of the magnetic mixing and the $\theta$-terms. 
We also considered the effects of
dyon and dyonic beads in the dark sector.

By considering SU(2)$_\mD$ behind the topological defects explicitly,
we clarified that the $\theta_\mD$-term affects the 
arguments only through the Witten effect.
We also found that the $\theta$-term of the QED sector plays no role in the absence of QED magnetic monopoles.

Magnetic and dyonic beads in the dark sector were found to have particularly interesting effects on QED coordination.
As found in Ref.\,\cite{Hiramatsu:2021kvu},
the kinetic mixing turns 
dark beads into pseudo-monopoles in the QED sector.
This result also applies to dark dyonic beads.
Besides, they
induce Coulomb potential for QED charges through the magnetic mixing,
which is not suppressed by $e^{-m_\mD'r}$ 
even in the $\uone[\mD]$ broken phase.
The dark electric charge of a dark dyon, on the other hand, only induces 
exponentially decaying electric potential for QED charges.

In this paper, we have focused on the ground states of a given topological charge in the dark sector. The phenomenological and cosmological implications are left for future work.

\begin{acknowledgements}
This work is supported by Grant-in-Aid for Scientific Research from the Ministry of Education, Culture, Sports, Science, and Technology (MEXT), Japan, 18H05542, 21H04471, 22K03615 (M.I.). 
This research is also supported by FoPM, WINGS Program, the University of Tokyo.
\end{acknowledgements}

\appendix

\section{Derivation of the Noether Charge}
\label{sec:noether}
In this appendix, we present the calculation of the Noether charge for the $\uone[\mD]$ global transformation \eqref{eq:U1D trans}.
The Noether charge is, in the temporal gauge,
\begin{align}
    N_{\uone[\mD]} &:= \int \dd[3]{x} \pdv{\lagrangian}{\pqty{\partial_0 A^a_{\mD i}}} \delta A_{\mD i}^{a}\ , \\
    &= \int \dd[3]{x} \pqty{
        -F_{\mD}^{a0i}
        + \epsilon \frac{\monohiggs^a}{v_1} F^{0i}
        - \frac{e_\mD^2 \theta_\mD}{8\pi^2} \tilde{F}_\mD^{a0i}
        - \frac{\tM}{8\pi^2}\frac{\monohiggs^a}{v_1}\tilde{F}^{0i}
    } \pqty{-\frac{1}{e_\mD v_1}D_i \monohiggs^a}\ .
\end{align}

The contribution from the kinetic term is
\begin{equation}
    \frac{1}{e_\mD v_1}\int \dd[3]{x} F_{\mD}^{a0i} D_i \monohiggs^a
    = \frac{1}{e_\mD}\int \dd[2]S_i \pqty{\frac{\monohiggs^a}{v_1} F_{\mD}^{a0i}}
       - \frac{1}{e_\mD v_1}\int \dd[3]{x} \monohiggs^a D_i F_{\mD}^{a0i}\ .
\end{equation}
The surface integral reduces to $\qcharge[\mD]/e_\mD$.
The integrand of the other term can be written as
\begin{align}
    \monohiggs^a D_i F_{\mD}^{a0i} &= \monohiggs^a D_\mu F_{\mD}^{a0\mu} \\
    \label{eq:noether1}
    &= \monohiggs^a \bqty{\epsilon D_\mu \pqty{\frac{\monohiggs^a}{v_1}F^{0\mu}} - \frac{\tM}{8\pi^2}D_\mu \pqty{\frac{\monohiggs^a}{v_1}\tilde{F}^{0\mu}}}\ ,
\end{align}
where we used the equation of motion for $A_{\mD\mu}^a$
\begin{equation}
    \label{eq:SU2EoM}
    D_\mu F_\mD^{a\mu\nu} - \epsilon D_\mu \pqty{\frac{\monohiggs^a}{v_1}F^{\mu\nu}} + \frac{\tM}{8\pi^2}D_\mu \pqty{\frac{\monohiggs^a}{v_1}\tilde{F}^{\mu\nu}} = e_\mD\epsilon^{abc}\monohiggs^b D^\nu \monohiggs^c\ .
\end{equation}

The contribution from the kinetic mixing term is
\begin{align}
     -\frac{\epsilon}{e_\mD v_1}\int \dd[3]{x} \frac{\monohiggs^a}{v_1} F^{0i}D_i \monohiggs^a
     &= -\frac{\epsilon}{e_\mD}\int \dd[2]{S}_i \pqty{\frac{\monohiggs^a\monohiggs^a}{v_1^2}F^{0i}}
     +\frac{\epsilon}{e_\mD v_1}\int \dd[3]{x} D_i \pqty{\frac{\monohiggs^a}{v_1} F^{0i}}\monohiggs^a\ .
\end{align}
The surface integral becomes $-\epsilon \qcharge/e_\mD$. The second term cancels the first term of \eqrefer{eq:noether1}.

The contribution from the $\theta_\mD$-term is
\begin{align}
        \frac{e_\mD \theta_\mD}{8\pi^2 v_1}\int \dd[3]{x}  \tilde{F}_\mD^{a0i}D_i\monohiggs^a
        &= \frac{e_\mD \theta_\mD}{8\pi^2 }\int \dd[2]{S}_i \pqty{\frac{\monohiggs^a}{v_1}\tilde{F}_\mD^{a0i}} - \frac{e_\mD \theta_\mD}{8\pi^2 v_1}\int \dd[3]{x}  \monohiggs^aD_i\tilde{F}_\mD^{a0i} \\
        &= -\frac{e_\mD \theta_\mD}{8\pi^2}Q^m_\mD\ ,
\end{align}
where we used the Bianchi identity at the second equality.

Similarly, the contribution from the magnetic mixing term is
\begin{equation}
    \frac{1}{e_\mD v_1}\frac{\tM}{8\pi^2}\int \dd[3]{x} \frac{\monohiggs^a}{v_1}\tilde{F}^{0i} D_i\monohiggs^a 
    = \frac{1}{e_\mD}\frac{\tM}{8\pi^2}\int \dd[2]{S}_i \pqty{\frac{\monohiggs^a\monohiggs^a}{v_1^2}\tilde{F}^{0i}}- \frac{1}{e_\mD v_1}\frac{\tM}{8\pi^2}\int \dd[3]{x} \monohiggs^a D_i \pqty{\frac{\monohiggs^a}{v_1} \tilde{F}^{0i}}\ .
\end{equation}
This time, the charge term vanishes as there is no QED magnetic monopole.
The second term cancels the second term of \eqrefer{eq:noether1}.

Putting all together, the Noether charge is found to be
\begin{equation}
     N_{\uone[\mD]}= \frac{1}{e_\mD}\qcharge[\mD] - \frac{\epsilon}{e_\mD}\qcharge - \frac{\theta_\mD e_\mD}{8\pi^2}Q^m_\mD \ .
\end{equation}

In the presence of dark or QED elementary charges, 
the Noether charge 
has additional contributions from them through $F_D^{0i}$ and $F^{0i}$.%
\footnote{In this work, we only consider massive test particles.
In the case of Dirac fermions, we take the phase convention so that the Dirac mass term is real valued.
For a discussion on the phase of the fermion mass term see Ref.~\cite{Callan:1982au}.}
Thus, in the case of an elementary dark charge, $\ncharge[\mD]$, we find
\begin{equation}
    \qcharge[\mD] = \frac{e_\mD\ncharge[\mD]}{1-\epsilon^2} \ , \quad
    \qcharge = \frac{\epsilon e_\mD\ncharge[\mD]}{1-\epsilon^2}\ ,
\end{equation}
and hence,
\begin{equation}
      N_{\text{U(1)}_\mD} = \ncharge[\mD] \ ,
\end{equation}
which is a half integer as we are considering SU(2)$_\mD$.
For a QED charge, $\ncharge$, on the other hand, 
\begin{equation}
    \qcharge[\mD] = \frac{\epsilon e\ncharge}{1-\epsilon^2} \ , \quad
    \qcharge = \frac{ e\ncharge}{1-\epsilon^2}\ ,
\end{equation}
and hence,
$N_{\text{U(1)}_\mD}=0$.

The Noether charge of QED is given by
\begin{align}
    N_\mathrm{QED}&:=\int \dd[3]{x} J_\mathrm{QED}^0 \\
    &= \frac{1}{e}Q_\text{QED}^e - \frac{\epsilon}{e} Q_{\mD}^e + 
    \frac{\tM}{8\pi^2} Q_\mD^m\ . \label{eq:QEDnoether}
\end{align}
\modifiedat{PRDv2}{
Here, we have used the equation of motion,
\begin{align}
 \partial_\mu F^{\mu\nu}  - \epsilon 
 \partial_\mu
\left( 
 \frac{\monohiggs^a}{v_1}F_\mD^{a\mu\nu}
\right)
+ \frac{\tM}{8\pi^2} \partial_\mu 
\left(\frac{\monohiggs^a}{v_1}\tilde{F}_\mD^{a\mu\nu}\right) = e J_\mathrm{QED}^\nu\ ,
\end{align}
 to replace the Noether current $J_{\mathrm{QED}}^\mu$ with the \modifiedat{PRDv3}{field} strengths.}
Thus, QED electric charges satisfy
$N_\mathrm{QED}=\ncharge$, while
dark elementary charges satisfy $N_\mathrm{QED} = 0$.
Dark monopoles and dark dyons also satisfy $N_\mathrm{QED} = 0$.
Thus, the mini-charges induced to the QED sector do not spoil the compactness of $\uone$.

\section{Defects Stability}
\label{sec:stability}
In this appendix, we argue that the 
the topological defects are stable even
in the presence of the mixing terms.
In general, the non-zero energy ground state of a topologically nontrivial sector is stable.

The dark monopole and 
the dark string are associated with the topological numbers $\pi_2(S^3/S^1)=\mathbb{Z}$,
$\pi_1(S^3/\mathbb{Z}_2)=\mathbb{Z}_2$, respectively.
Thus, to ensure their stability, it suffices to show that they cannot reach energy zero.

Let us first consider
the dark monopole/dyon solutions. For the energy not to diverge, we need
\begin{align}
    F_\mD^{a\mu\nu} &= \order{r^{-2}}\ , \\
    \label{eq:Dphi}
    D_{\mu}\phi^a & = \order{r^{-2}}\ , \\
    F^{\mu\nu} &= \order{r^{-2}}\ ,
\end{align}
at $r\to\infty$.
Then, from \eqrefer{eq:Dphi}, 
we find that 
the magnetic charge is proportional to the topological number $n_{\mD}^m$,
\begin{align}
    Q_\mD^m = \int_{r\to \infty}
    \dd[2]S_i B_\mD^i =- \frac{1}{2e_\mD^2v_1^3}
    \int_{r\to \infty} \dd[2]S_i \epsilon_{ijk}\epsilon^{abc}\phi^a\partial_{j}\phi^b \partial_{k}\phi^c\  .
\end{align}
Thus, the solutions with non-trivial topological number are associated with the non-vanishing magnetic field, and hence, they have non-vanishing energy.
Thus, such solutions (i.e. the local minimum of the energy) 
with non-trivial topological number are stable.
The mixing terms do not modify this argument.

In the case of the dark string,
non-divergent tension requires $D_\mu \chi = 0$ at $\rho \to \infty$.
In this case, the cosmic strings 
with non-trivial winding number 
have non-vanishing magnetic flux along them.
Thus, the tension of the cosmic strings is non-vanishing.
Again, the mixing terms are irrelevant here.

Finally, let us discuss the stability of the bead solution.
As we assume hierarchical VEVs between $\monohiggs$ and $\strhiggs$, the topological arguments of the monopole/dyon are 
not affected by the cosmic strings attached to them.
Since the stability of the monopole/dyon are not affected by the mixing terms, they do not spoil the stability of the bead solution either.

\bibliography{Refs.bib}
\end{document}